\documentclass[a4paper]{article}

\usepackage{INTERSPEECH2022}

\usepackage{graphicx}
\usepackage{amsmath}
\usepackage{amssymb}
\usepackage{booktabs}
\usepackage{multirow}
\usepackage{diagbox}
\usepackage[hang,flushmargin]{footmisc}
\usepackage[hyphens]{url}

\usepackage{hyperref}
\hypersetup{
colorlinks=true,
linkcolor=blue,
citecolor=blue,
urlcolor=blue,
}

\usepackage{colortbl}
\definecolor{lightgray}{gray}{0.95}

\title{Disentangling the Impacts of Language and Channel Variability on Speech Separation Networks}
\name{Fan-Lin Wang$^1$, Hung-Shin Lee$^1$, Yu Tsao$^2$, and Hsin-Min Wang$^1$}
\address{
$^1$Institute of Information Science, Academia Sinica\\
$^2$Research Center for Information Technology Innovation, Academia Sinica}
\email{\{faliwang1999,hungshinlee\}@gmail.com}

\begin{document}

\maketitle
\begin{abstract}
Because the performance of speech separation is excellent for speech in which two speakers completely overlap, research attention has been shifted to dealing with more realistic scenarios. However, domain mismatch between training/test situations due to factors, such as speaker, content, channel, and environment, remains a severe problem for speech separation. Speaker and environment mismatches have been studied in the existing literature. Nevertheless, there are few studies on speech content and channel mismatches. Moreover, the impacts of language and channel in these studies are mostly tangled. In this study, we create several datasets for various experiments. The results show that the impacts of different languages are small enough to be ignored compared to the impacts of different channels. In our experiments, training on data recorded by Android phones leads to the best generalizability. Moreover, we provide a new solution for channel mismatch by evaluating projection, where the channel similarity can be measured and used to effectively select additional training data to improve the performance of in-the-wild test data.

\end{abstract}
\noindent\textbf{Index Terms}: speech separation, domain mismatch

\section{Introduction}
The cocktail party problem \cite{Haykin2005} refers to the perception of each speech source in a noisy social environment, and has been extensively studied for many years. As the necessary pre-processing for downstream tasks, such as speaker diarization \cite{Neumann2019} and automatic speech recognition \cite{Raj2020a}, many efforts have been made in speech separation \cite{Wang2018} to address this problem.

Nowadays, most of the studies on speech separation are conducted on the WSJ0-2mix dataset \cite{Hershey2016}. WSJ0-2mix was artificially synthesized from WSJ0\footnote{\url{https://catalog.ldc.upenn.edu/LDC93S6A}}, an English speech corpus. In WSJ0-2mix, all mixed utterances are full overlaps of clean speech of two speakers. A typical architecture among the popular methods today is the time-domain audio separation network (TasNet) \cite{Luo2018}. Many works based on TasNet have achieved extraordinary performance \cite{Luo2020,Chen2020,Zeghidour2020,Hu2021,Subakan2021}. However, WSJ0-2mix unrealistically sets a limit on the fixed number of speakers (addressed by WSJ0-\textit{n}mix), the fully overlapping ratio, and the quiet scenario. To push the field towards more practical and practicable scenarios, many efforts have been made.

Some research focused on the agnosticism of the number of speakers in utterances. In \cite{kinoshita2018,Takahashi2019}, the model extracts the speakers recursively. In \cite{Zhu2021}, the authors used a counter head to determine the number of speakers. Still others tried to extract the target speakers by some sort of information, such as visual features \cite{Chung2020}, speech recognition information \cite{Erdogan2015}, and anchor audio clips \cite{Han2020}. In addition, some expanded the studies by evaluating different ratios of overlapping speech. For example, a sparsely overlapping version of WSJ0-2mix was introduced in \cite{Menne2019}, and the study found the challenges of Deep Clustering \cite{Hershey2016} on such mixtures. In order to simulate the actual meeting, LibriCSS \cite{ChenLibri2020}, a test set containing a lower overlapping ratio and longer speech utterances, was made suitable for research focusing on downstream speech recognition tasks \cite{Li2021,Li2021a}. 

Despite the broad aspects of these previous works, domain mismatch between training/test situations remains challenging for the application in actual scenarios, since all of the datasets are synthesized artificially from audios recorded in anechoic chambers. Domain mismatch can be attributed to four factors: speaker, content, channel, and environment. With regards to speaker mismatch, speakers in the test sets of all datasets are unseen in training sets. However, the performance does not have an apparent drop, thus demonstrating the speaker generalizability of the models. As for the environment mismatch, we refer to the reverberation and noises we may encounter in the reality. To deal with the issue, two datasets have been presented: WHAM! \cite{Wichern2019} and WHAMR! \cite{Maclejewski2020}, which are the noisy and reverberant extensions of WSJ0-2mix, respectively. Furthermore, the authors of \cite{Subakan2021a} presented a realistic dataset that is recorded by multiple speakers in the same reverberant chamber. In \cite{Petermann2021}, the authors presented a new task for separating speech, music, and sound effects.

In particular, content and channel mismatch are rarely investigated. Content mismatch focuses on the contents spoken by speakers. In \cite{Kadioglu2020} and \cite{cosentino2020}, they believed that the larger the vocabulary presented in the training set, the better generalizability the model possesses. Moreover, content mismatch is not restricted by the vocabulary. It can also be extended to different languages containing various phonemes. In \cite{Borsdorf2021}, the GlobalPhoneMS2 dataset consisting of 22 spoken languages was presented, and the authors showed that when trained on a multilingual dataset, the networks can improve their performance on unseen languages as well as nearly all seen languages. However, the dataset can only entangle the languages with the recording characteristics, which means that we can not determine which aspect dominates the performance. Regarding channel mismatch, it focuses on the type of microphones used in the recording. The authors of \cite{Maciejewski2019} considered that near-field data is easier to separate than far-field data, although both are recorded in the same environment.  In the pandemic era, virtual meetings become prevalent, and a wider variety of microphones are used for recording. Therefore, channel mismatch should be investigated in more depth to meet demand.

To delve into different aspects, we leverage several corpora and synthesize them to create datasets of mixtures with similar statistics to WSJ0-2mix. Our experiments are mainly divided into two categories. First, we focus on the impact of language on some representative separation models. Second, we control the language factors and compare them across different channels. Details and source codes are open-sourced\footnote{\url{https://github.com/Sinica-SLAM/COSPRO-mix}}.

\section{Datasets for Speech Separation}
\label{sec:dataset}
All the datasets used in this study (except WSJ0-2mix, which is widely used in speech separation papers) are introduced in this section, including which task the dataset is used for and how the dataset is used. Among these datasets, COSPRO \cite{Tseng2003} and TAT \cite{Liao2020} will be discussed in more detail, as these two datasets were used as training sets in our experiments and were not used in other speech separation papers.

\subsection{COSPRO}
\label{ssec:cospro}
Mandarin Continuous Speech Prosody Corpora (COSPRO) \cite{Tseng2003} was designed, collected, and annotated at the Institute of Linguistics, Academia Sinica. As the name suggests, the main feature of this dataset is continuous or fluent speech prosody and its influence on speech production and perception. COSPRO includes a total of approximately 130 hours of read speech and only 0.3 hours of spontaneous narration in Mandarin Chinese. There are nine sets of speech corpora, summarized in Table \ref{tab:cospro}. Each corpus was designed to bring out different prosodic features involved in fluent speech. 

\subsection{COSPRO-2mix}
\label{ssec:cospro-2mix}
Before mixing the utterances in COSPRO, we first removed files that were inappropriate, such as utterances containing only a few characters or carrier sentences that differ a lot from actual conversations. Because most utterances are much longer than those in WSJ0, we used a pre-trained speech recognizer to segment the utterances. To this end, forced alignment was performed on all utterances with word-level transcriptions using the Kaldi toolkit \cite{Povey2011}. A time-marked conversation file (CTM) containing time-aligned word transcriptions was obtained for each utterance. Then, we trimmed each utterance into multiple consecutive segments based on its CTM, subject to the limit of the maximum segment length (approximately 10 seconds). Every cut point must be at a word boundary.

After segmentation, we referred to how WSJ0-2mix was mixed to make a dataset comparable to WSJ0-2mix. Our procedure for making COSPRO-2mix is as follows. First, we selected 81 speakers (34 males and 47 females) from the training set, each with more than 400 seconds of speech. In COSPRO, there is a dataset consisting of more speakers, but some of them with fewer utterances. To avoid overfitting in training, we only selected speakers with longer speaking times. For the evaluation set, seven males and seven females were selected. Then, we followed the mixing procedure of WSJ0-2mix. For each mixed utterance in WSJ0-2mix, we constructed a corresponding mixed utterance based on the same gender combination and volume ratio. In this way, COSPRO-2mix and WSJ0-2mix have the same number of mixed utterances and similar statistics. COSPRO-2mix was used in language-related experiments.

\begin{table}[t]
\caption{Information about the nine corpora in COSPRO.}
\vspace{-5pt}
\centering
\label{tab:cospro}
\begin{tabular}{c@{\hspace{5pt}}l@{\hspace{5pt}}c@{\hspace{5pt}}c}
\toprule
\multirow{2}{*}{\textbf{No.}} & \multirow{2}{*}{\textbf{Feature}} & \multirow{2}{*}{\shortstack[1]{\# Speakers\\(M / F)}} & \multirow{2}{*}{\shortstack[1]{Length\\(hours)}} \\
&&&\\
\midrule
\midrule
01 & phonetically-balanced & 3 / 3 & 18.63 \\
\rowcolor{lightgray} 02 & multiple-speaker & 40 / 50 & 19.48 \\
03 & intonation-balanced & 2 / 3 & 31.16 \\
\rowcolor{lightgray} 04 & stress-pattern balanced & 1 / 1 & 0.80 \\
05 & lexically-balanced & 1 / 1 & 35.83 \\
\rowcolor{lightgray} 06 & focus-balanced prosody groups & 1 / 1 & 7.50 \\
07 & multiple text-type/speaking-style & 1 / 1 & 1.53 \\
\rowcolor{lightgray} 08 & prosody-unit balanced & 1 / 1 & 15.20 \\
09 & comparable spontaneous/read & 1 / 1 & 0.70 \\
\bottomrule
\end{tabular}
\vspace{-15pt}
\end{table}

\subsection{English Across Taiwan (EAT) \& EAT-2mix}
\label{ssec:eat}
The English Across Taiwan (EAT) corpus\footnote{\url{http://www.aclclp.org.tw/doc/eat_brief_e.pdf}} contains microphone and telephone recordings of college students in Taiwan. We selected English utterances and English/Mandarin code-mixing utterances from the microphone recordings to avoid the channel mismatch as much as possible and created an English test set ($\textnormal{EAT-2mix}_\textnormal{Eng}$) and an English/Mandarin code-mixing test set ($\textnormal{EAT-2mix}_\textnormal{E\&M}$) in the same way as COSPRO-2mix. Both test sets contain the same speakers and similar utterance lengths. With this dataset, we can compare performance within the same channel, which means that the performance gap is mainly due to differences in content. The two test sets were used in the first part of our experiment.

\subsection{Taiwanese Across Taiwan (TAT)}
The Taiwanese Across Taiwan (TAT) corpus \cite{Liao2020} was originally intended for Taiwanese speech recognition. Taiwanese, also known as Southern Min, is a common dialect in Taiwan and belongs to the same language family as Mandarin. The 200 speakers contributing to the corpus include men, women, and children of all ages across Taiwan, with content ranging from written articles to everyday conversations. Most notably, the corpus was recorded simultaneously on six channels, including different microphones and smartphones. This feature helps us investigate the impacts of different channels. The way of synthesizing the datasets of mixed utterances is described below.

\subsection{TAT-2mix}
\label{ssec:TAT-mix}
The six channels in TAT include: one close-talk (Audio-Technica AT2020), one distant X-Y stereo microphone (ZOOM XYH-6 stereo microphone, containing left and right channels), one lavalier (Superlux WO518+PS418D), iOS devices (including iPhones, iPads, and iPods), and Android phones (produced by ASUS and Samsung). According to these six channels, we created six datasets respectively: $\textnormal{TAT-2mix}_\textnormal{condenser}$, $\textnormal{TAT-2mix}_\textnormal{XYH-6-X}$, $\textnormal{TAT-2mix}_\textnormal{XYH-6-Y}$, $\textnormal{TAT-2mix}_\textnormal{lavalier}$, $\textnormal{TAT-2mix}_\textnormal{iOS}$, and $\textnormal{TAT-2mix}_\textnormal{Android}$.

Before mixing the utterances in TAT, we segmented the utterances using the same method as in COSPRO-2mix to trim the noise in the front and tail of the audio. Since the original length was already appropriate for speech separation tasks, we did not cut the recording into shorter utterances.

After segmentation, the mixing procedure is similar to that of WSJ0-2mix. We randomly selected 101 speakers (49 males and 52 females) for training and validation. On the other hand, 18 speakers (11 males and 7 females) were selected for testing. The number of speakers and the gender distribution are exactly the same as WSJ0-2mix. Then, we followed the mixing procedure of WSJ0-2mix to assure that TAT-2mix and WSJ0-2mix have the same number of mixed utterances and similar statistics. All datasets are comprised of the same mixtures of utterances but from different channels.

\section{Speech Separation Models}
\label{sec:model}
In our experiments, we evaluated three representative speech separation models, namely, Conv-TasNet \cite{Luo2019}, DPRNN \cite{Luo2020}, and DPT-Net \cite{Chen2020}. These models have been implemented in the Asteroid toolkit \cite{Pariente2020} so that it is easy to compare the three models in the same coding environment. All three models are based on TasNet consisting of an encoder, a separator, and a decoder. The encoder is mainly a 1D convolutional layer, which transforms a waveform into an embedding space. On the other hand, the decoder is a 1D transposed convolutional layer, which transforms the embedding back to the waveform. The main difference between the three models is the structure of the separator. 

In Conv-TasNet, the separator is a fully-convolutional separation module, which is composed of stacked 1D dilated convolutional blocks with increasing dilation factors. The dilation factors increase exponentially to ensure that a temporal context window is sufficiently large to take advantage of the long-range dependency of the speech signal. In DPRNN, the separator splits the output of the encoder into chunks with or without overlap, concatenates them to form a 3D tensor, and then passes it to a dual-path BiLSTM module for processing in two different dimensions: the chunk size and the number of chunks. The two paths can help RNN see the information around and far away from the current time frame to achieve better performance. As for DPT-Net, the idea is similar to DPRNN. However, DPT-Net replaces the BiLSTM module with a Transformer encoder, and replaces the first fully connected layer with an RNN in the feed-forward network to learn the order information. It has shown the best performance among these three models.

In our experiments, all model configurations followed the original configuration implemented in Asteroid, except that we applied six dual-path Transformer blocks in DPT-Net, which conforms to the configuration in the original paper. We used the same loss function in terms of the scale-invariant signal-to-distortion ratio (SI-SNR). $\text{SI-SNR}(\mathbf{x})$ is defined by
\begin{equation}
\label{eq:si-snr}
10\log_{10}\frac{\langle\mathbf{\tilde{x}},\mathbf{\tilde{x}}\rangle}{\langle \mathbf{e},\mathbf{e}\rangle}\quad(\mathbf{\tilde{x}}=\frac{\langle\mathbf{x},\mathbf{s}\rangle}{\langle\mathbf{x},\mathbf{x}\rangle} \mathbf{x}\quad\textrm{and}\quad\mathbf{e}=\mathbf{\tilde{x}}-\mathbf{s}),
\end{equation}
where $\mathbf{x}$ denotes the output of the network and should ideally be equal to the clean source $\mathbf{s}$. We only trained 100 epochs for each model to speed up the progress of the experiments.

\section{Experiments}
\label{sec:experiment}

\subsection{Impact of Language}
\label{ssec:language}

We trained the representative models on WSJ0-2mix and COSPRO-2mix and evaluated them on several datasets. The results are shown in Table \ref{tab:language}, where the columns refer to the training sets and the separation models; the rows refer to the test sets. The metric used in this study is SI-SNRi (dB), the improvement of SI-SNR (cf. Eq. \ref{eq:si-snr}), defined by
\begin{equation}
\label{eq:si-snri}
\textrm{SI-SNRi}=\textrm{SI-SNR}(\mathbf{x})-\textrm{SI-SNR}(\mathbf{m}),
\end{equation}
where $\mathbf{m}$ and $\mathbf{x}$ denote the input mixture and output of the network, respectively.

First, comparing each column in Table \ref{tab:language} under the two training sets, the performance rankings are in line with past results: DPT-Net outperforms DPRNN, and DPRNN outperforms Conv-TasNet. Then, we examine the performance in each row. Looking at the first three rows (the group of WSJ0 \& COSPRO), we can see that the results of WSJ0-2mix and COSPRO-2mix are comparable, under both in-domain (training/testing from the same dataset) and cross-domain (training/testing from the different dataset) testing conditions. In WSJ0$\times$COSPRO, each utterance is a mixture of a Mandarin utterance from COSPRO and an English utterance from WSJ0. Due to different languages and different recording conditions, WSJ0$\times$COSPRO is easier to separate than WSJ0-2mix and COSPRO-2mix.

The results in the fourth and fifth rows of Table \ref{tab:language} (the group of EAT) show the impact of language more explicitly since the only difference between the two test sets is the language. We observe that the performance on EAT is much worse than those on COSPRO-2mix and WSJ0-2mix, because the clean waveform contains some microphone noise, which may degrade SI-SNR, and the recording quality is not high, which causes many pop and explosive sounds. Compared to the performance in other rows, it is obvious that the performance gap between $\textnormal{EAT-2mix}_\textnormal{E\&M}$ and $\textnormal{EAT-2mix}_\textnormal{Eng}$ is almost negligible. That is, although language has an impact on speech separation, the impact is so small to neglect, complying with the results in \cite{Healy2021}.

The results in the remaining rows of Table \ref{tab:language} (the group of TAT) reveal the impact of channels since the only difference between these test sets is the channel. There is a large difference between the performance of each row, suggesting that channel variability seems to dominate the performance. We also found that when the test set is more difficult, such as $\textnormal{TAT-2mix}_\textnormal{iOS}$, the performance difference between models is less pronounced. To investigate the impact of channel variability and the relationship between channels, we performed the next set of experiments.

\begin{table}[t]
\caption{Performance (SI-SNRi in dB) of various separation models (\textbf{ConvT}: Conv-TasNet, \textbf{DPR}: DPRNN, \textbf{DPT}: DPT-Net) tested on various datasets and trained on WSJ0-2mix and COSPRO-2mix. WSJ0$\times$COSPRO is a test set, where each utterance is a mixture of a Mandarin utterance from COSPRO and an English utterance from WSJ0.}
\vspace{-5pt}
\centering
\label{tab:language}
\begin{tabular}{l@{\hspace{4pt}}c@{\hspace{4pt}}c@{\hspace{4pt}}c@{\hspace{4pt}}c@{\hspace{4pt}}c@{\hspace{4pt}}c}
\toprule
\multirow{2}{*}{\diagbox{\textbf{Test}}{\textbf{Train}}} & \multicolumn{3}{c}{\textbf{WSJ0-2mix}} & \multicolumn{3}{c}{\textbf{COSPRO-2mix}} \\
\cmidrule{2-4}
\cmidrule{5-7}
& \textbf{ConvT} & \textbf{DPR} & \textbf{DPT} & \textbf{ConvT} & \textbf{DPR} & \textbf{DPT} \\
\midrule
\midrule
WSJ0-2mix & 15.84 & 16.65 & 17.11 & 12.49 & 13.15 & 13.81 \\
COSRPO-2mix & 12.17 & 13.19 & 14.78 & 15.91 & 16.92 & 17.23\\
WSJ0$\times$COSPRO & 16.09 & 16.95 & 18.14 & 16.33 & 16.88 & 17.41 \\ 
\midrule
$\textnormal{EAT-2mix}_\textnormal{E\&M}$ & 3.02 & 3.93 & 4.77 & 2.68 & 3.16 & 3.20 \\
$\textnormal{EAT-2mix}_\textnormal{Eng}$ & 2.84 & 3.90 & 4.46 & 2.42 & 2.83 & 2.89 \\
\midrule
$\textnormal{TAT-2mix}_\textnormal{Android}$ & 2.25 & 3.15 & 4.35 & 3.94 & 4.08 & 4.70 \\
\rowcolor{lightgray} $\textnormal{TAT-2mix}_\textnormal{iOS}$ & -0.44 & 0.68 & 0.78 & 0.94 & 0.87 & 1.86 \\
$\textnormal{TAT-2mix}_\textnormal{condenser}$ & 6.99 & 7.42 & 9.23 & 2.32 & 3.21 & 3.44 \\
\rowcolor{lightgray} $\textnormal{TAT-2mix}_\textnormal{lavalier}$ & 5.39 & 6.55 & 7.57 & 5.91 & 3.75 & 5.96\\
$\textnormal{TAT-2mix}_\textnormal{XYH-6-X}$ & 0.64 & 1.45 & 1.51 & -0.66 & 0.51 & 0.75 \\
\rowcolor{lightgray} $\textnormal{TAT-2mix}_\textnormal{XYH-6-Y}$ & -0.10 & 0.61 & 0.99 & -0.60 & -0.05 & 0.68 \\
\bottomrule
\end{tabular}
\vspace{-15pt}
\end{table}

\subsection{Impact of Channel Variability}
\label{ssec:channel}

In this set of experiments, we trained a DPT-Net on each TAT-2mix recorded by a specific channel and evaluated on all TAT-2mix separately. The results are shown in Table \ref{tab:channel}. The columns stand for the training sets, and the rows present the test sets.

Overall, we can again see the severity of the impact of channel variability. Although all datasets are composed of the same speech content and speakers, the performance varies. In addition, the models perform best when the test data is in-domain (training and testing are from the same channel), in line with previous studies that emphasized the importance of in-domain training data \cite{Kadioglu2020,cosentino2020,sivaraman2021}.

If we examine the SI-SNRi values in Table \ref{tab:channel}, \textbf{condenser} and \textbf{lavalier} are the easiest channels, and \textbf{XYH-6-X} and \textbf{XYH-6-Y} are the hardest. The result is reasonable because \textbf{condenser} is closest to the speaker (5--10cm) with a pop filter, and \textbf{lavalier} is also very close to the speaker's mouth. \textbf{Android} and \textbf{iOS} are 15--20cm from the speaker. The distance between \textbf{XYH-6} and the speaker is about one meter. As reported in \cite{Maciejewski2019}, speech separation is more difficult for distant microphones. 

By comparing the columns of \textbf{XYH-6-X} and \textbf{XYH-6-Y}, it is clear that the performance between the left and right channels is almost equivalent, but \textbf{XYH-6-X} has a slight advantage. When comparing the rows of \textbf{XYH-6-X} and \textbf{XYH-6-Y}, \textbf{XYH-6-Y} is slightly harder than \textbf{XYH-6-X}. As expected, the left and right channels are somewhat similar. As for the two smartphone systems, whether comparing the rows of \textbf{Android} and \textbf{iOS} to see how other models perform on the two test sets, or comparing the columns of \textbf{Android} and \textbf{iOS} to examine the generalizability of the two models, the results show that the characteristics of the iOS microphones and the Android microphones differ considerably. However, for these two smartphone systems, their counterpart model still performs the best among all cross-domain models, e.g., the \textbf{iOS} model gives an SI-SNRi of 7.60 when tested on the Android test set, which is better than the SI-SNRi values of other cross-domain models.

Furthermore, the model trained on $\textnormal{TAT-2mix}_\textnormal{Android}$ achieves the highest average performance on all test sets, indicating its robustness. One possible reason is that \textbf{Android} contains smartphones from different companies, with different frequency responses for their built-in microphones. It is also worth mentioning that the model trained on $\textnormal{TAT-2mix}_\textnormal{condenser}$ does not generalize well to other channels. However, condenser microphones are the most popular devices for recording corpora, which poses a problem in real-world scenarios. In  the next subsection, we proposed a solution that effectively selects additional training data to improve performance on the in-the-wild test data.

\begin{table}[t]
\caption{Cross-dataset evaluation (SI-SNRi in dB) of TAT-2mix on various channels. The model for evaluation is DPT-Net.}
\vspace{-5pt}
\centering
\label{tab:channel}
\begin{tabular}{l@{\hspace{6pt}}c@{\hspace{6pt}}c@{\hspace{6pt}}c@{\hspace{6pt}}c@{\hspace{6pt}}c@{\hspace{6pt}}c}
\toprule
\multirow{2}{*}{\diagbox{\textbf{Test}}{\textbf{Train}}} & \multirow{2}{*}{\shortstack[1]{\textbf{And-}\\\textbf{roid}}} &
\multirow{2}{*}{\textbf{iOS}} & 
\multirow{2}{*}{\shortstack[1]{\textbf{cond-}\\\textbf{enser}}} & 
\multirow{2}{*}{\shortstack[1]{\textbf{lav-}\\\textbf{alier}}} & 
\multirow{2}{*}{\shortstack[1]{\textbf{XYH-}\\\textbf{6-X}}} & 
\multirow{2}{*}{\shortstack[1]{\textbf{XYH-}\\\textbf{6-Y}}} \\
&&&&&& \\
\midrule
\midrule
Android & \textbf{10.89} & 7.60 & 5.32 & 6.98 & 6.19 & 5.59 \\
\rowcolor{lightgray} iOS & 8.42 & \textbf{10.69} & 1.33 & 3.97 & 4.10 & 4.07 \\
condenser & 10.47 & 4.53 & \textbf{14.07} & 11.97 & 9.95 & 9.26 \\
\rowcolor{lightgray} lavalier & 11.62 & 9.57 & 9.34 & \textbf{14.10} & 5.62 & 4.24 \\
XYH-6-X & 5.87 & 3.10 & 6.83 & 3.10 & \textbf{8.82} & 8.03 \\
\rowcolor{lightgray} XYH-6-Y & 4.65 & 2.51 & 5.44 & 1.86 & 7.34 & \textbf{8.15} \\
\midrule
Average & 8.65 & 6.33 & 7.06 & 7.00 & 7.00 & 6.56  \\
\bottomrule
\end{tabular}
\vspace{-15pt}
\end{table}

\subsection{Data Selection by Pairwise PCA}
\label{ssec:p-pca}

We performed pairwise principal component analysis (PCA) to visualize the relationship between channels. First, we encoded the results of individual test sets (i.e., the rows in Table \ref{tab:channel}) into row vectors $\{\mathbf{v}_i\}$. Then, we constructed the covariance matrix $\Lambda$ by 
\begin{equation} 
\label{eq:covariance}
\Lambda = \frac{1}{2N} \sum_{i=1}^{N} \sum_{j=1}^{N}\left(\mathbf{v}_i-\mathbf{v}_j\right)^\intercal\left(\mathbf{v}_i-\mathbf{v}_j\right),
\end{equation}
where $N$ denotes the number of vectors. We performed eigen-decomposition on $\Lambda$ to find the two eigenvectors with the largest eigenvalues as the two directions of the new feature space. Then, $\{\mathbf{v}_i\}$ was projected onto the new 2D feature space.

\begin{figure}[t]
\begin{center}
\includegraphics[width=0.47\textwidth]{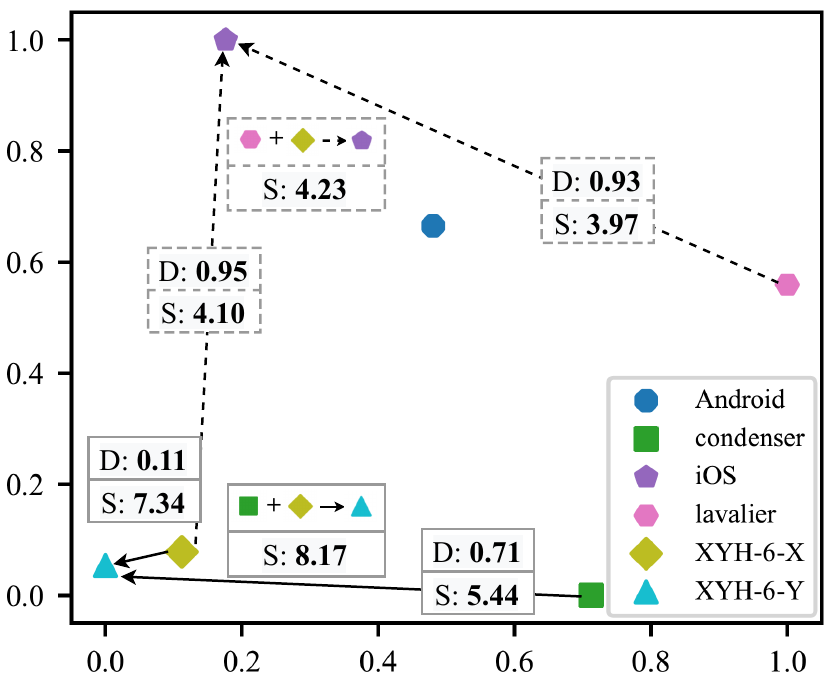}
\end{center}
\vspace{-15pt}
\caption{2D projection of the TAT-2mix test sets using pairwise PCA. D denotes the distance between two points, and S denotes the SI-SNRi (dB) by the model trained on the dataset at the arrow tail and tested on the dataset at the arrow tip.}
\label{fig:p-pca}
\vspace{-15pt}
\end{figure}

Fig. \ref{fig:p-pca} shows the scatterness of the projected points. From the figure, we can see that \textbf{XYH-6-X} and \textbf{XYH-6-Y} are close. In order to evaluate the usefulness of distance, we conducted two experiments. We found that if the training set contains data close to the test set, the performance improves significantly. For example, as shown by the solid arrows, when the training set contains $\textnormal{TAT-2mix}_\textnormal{condenser}$ and $\textnormal{TAT-2mix}_\textnormal{XYH-6-X}$, the performance on $\textnormal{TAT-2mix}_\textnormal{XYH-6-Y}$ can be improved by at least 11\% compared to the performance of the model trained on $\textnormal{TAT-2mix}_\textnormal{XYH-6-X}$ (8.17 vs 7.34). In contrast, if both training sets are far away from the test data, even doubling the size of the training data may not improve the performance much. For example, the dotted arrows show that the performance on $\textnormal{TAT-2mix}_\textnormal{iOS}$ is only improved by at most 7\% comparing the model trained on $\textnormal{TAT-2mix}_\textnormal{XYH-6-X}$ \& $\textnormal{TAT-2mix}_\textnormal{lavalier}$ and the model trained on $\textnormal{TAT-2mix}_\textnormal{lavalier}$ (4.23 vs 3.97).

The above results suggest that additional training data should be chosen wisely to effectively enhance the robustness of the model. Therefore, we propose the following solution. When encountering new data recorded by an unknown channel, we can first use these six models to perform speech separation on the data. Then, the test results are projected onto the 2D space, so that we can observe the distances between the new dataset and other points. Finally, we select existing datasets close to the new point and add them to the training data to improve the performance of the model on the new data.

\section{Conclusions and Future Work}
\label{sec:conclusion}

In this paper, we have investigated the impact of language and channel variability on speech separation by designing proportionate datasets and comparing cross-domain performance. In the first set of experiments, we created COSPRO-2mix and found that the impact of language is negligibly small compared to channel variability. In the second set of experiments, we created TAT-2mix from TAT recorded by different channels to observe the relationship between the channels. Moreover, we proposed a solution to address channel mismatch by evaluating projection and effectively selecting training data. In the future, we hope to prove our method on in-the-wild test data and extend it to domain generalization, where evaluation data are unknown.

\newpage
\bibliographystyle{IEEEtran}
\bibliography{references.bib}

\begin{thebibliography}{10}
\providecommand{\url}[1]{#1}
\csname url@samestyle\endcsname
\providecommand{\newblock}{\relax}
\providecommand{\bibinfo}[2]{#2}
\providecommand{\BIBentrySTDinterwordspacing}{\spaceskip=0pt\relax}
\providecommand{\BIBentryALTinterwordstretchfactor}{4}
\providecommand{\BIBentryALTinterwordspacing}{\spaceskip=\fontdimen2\font plus
\BIBentryALTinterwordstretchfactor\fontdimen3\font minus
  \fontdimen4\font\relax}
\providecommand{\BIBforeignlanguage}[2]{{%
\expandafter\ifx\csname l@#1\endcsname\relax
\typeout{** WARNING: IEEEtran.bst: No hyphenation pattern has been}%
\typeout{** loaded for the language `#1'. Using the pattern for}%
\typeout{** the default language instead.}%
\else
\language=\csname l@#1\endcsname
\fi
#2}}
\providecommand{\BIBdecl}{\relax}
\BIBdecl

\bibitem{Haykin2005}
S.~Haykin and Z.~Chen, ``{The cocktail party problem},'' \emph{Neural
  Computation}, vol.~17, pp. 1875--1902, 2005.

\bibitem{Neumann2019}
T.~V. Neumann, K.~Kinoshita, M.~Delcroix, S.~Araki, T.~Nakatani, and
  R.~Haeb-Umbach, ``{All-neural online source separation, counting, and
  diarization for meeting analysis},'' in \emph{Proc. ICASSP}, 2019.

\bibitem{Raj2020a}
D.~Raj, P.~Denisov, Z.~Chen, H.~Erdogan, Z.~Huang, M.~He, S.~Watanabe, J.~Du,
  T.~Yoshioka, Y.~Luo, N.~Kanda, J.~Li, S.~Wisdom, and J.~R. Hershey,
  ``{Integration of speech separation, diarization, and recognition for
  multi-speaker meetings: System description, comparison, and analysis},'' in
  \emph{Proc. IEEE SLT}, 2021.

\bibitem{Wang2018}
D.~Wang and J.~Chen, ``{Supervised speech separation based on deep learning: An
  overview},'' \emph{IEEE/ACM Transactions on Audio Speech and Language
  Processing}, vol.~26, no.~10, pp. 1702--1726, 2018.

\bibitem{Hershey2016}
J.~R. Hershey, Z.~Chen, J.~Le~Roux, and S.~Watanabe, ``{Deep clustering:
  Discriminative embeddings for segmentation and separation},'' in \emph{Proc.
  ICASSP}, 2016.

\bibitem{Luo2018}
Y.~Luo and N.~Mesgarani, ``{TasNet: Time-domain audio separation network for
  real-time, single-channel speech separation},'' in \emph{Proc. ICASSP}, 2018.

\bibitem{Luo2020}
Y.~Luo, Z.~Chen, and T.~Yoshioka, ``{Dual-path RNN: Efficient long sequence
  modeling for time-domain single-channel speech separation},'' in \emph{Proc.
  ICASSP}, 2020.

\bibitem{Chen2020}
J.~Chen, Q.~Mao, and D.~Liu, ``{Dual-path transformer network : Direct
  context-aware modeling for end-to-end monaural speech separation},'' in
  \emph{Proc. Interspeech}, 2020.

\bibitem{Zeghidour2020}
\BIBentryALTinterwordspacing
N.~Zeghidour and D.~Grangier, ``{Wavesplit: End-to-end speech separation by
  speaker clustering},'' 2020. [Online]. Available:
  \url{https://arxiv.org/abs/2002.08933}
\BIBentrySTDinterwordspacing

\bibitem{Hu2021}
X.~Hu, K.~Li, W.~Zhang, Y.~Luo, J.-M. Lemercier, and T.~Gerkmann, ``{Speech
  separation using an asynchronous fully recurrent convolutional neural
  network},'' in \emph{Proc. NeurIPS}, 2021.

\bibitem{Subakan2021}
C.~Subakan, M.~Ravanelli, S.~Cornell, M.~Bronzi, and J.~Zhong, ``{Attention is
  all you need in speech separation},'' in \emph{Proc. ICASSP}, 2021.

\bibitem{kinoshita2018}
K.~Kinoshita, L.~Drude, M.~Delcroix, and T.~Nakatani, ``{Listening to each
  speaker one by one with recurrent selective hearing networks},'' in
  \emph{Proc. ICASSP}, 2018.

\bibitem{Takahashi2019}
N.~Takahashi, S.~Parthasaarathy, N.~Goswami, and Y.~Mitsufuji, ``{Recursive
  speech separation for unknown number of speakers},'' in \emph{Proc.
  Interspeech}, 2019.

\bibitem{Zhu2021}
J.~Zhu, R.~Yeh, and M.~Hasegawa-Johnson, ``{Multi-decoder DPRNN: High accuracy
  source counting and separation},'' in \emph{Proc. ICASSP}, 2021.

\bibitem{Chung2020}
S.~W. Chung, S.~Choe, J.~S. Chung, and H.~G. Kang, ``{FaceFilter: Audio-visual
  speech separation using still images},'' in \emph{Proc. Interspeech}, 2020.

\bibitem{Erdogan2015}
H.~Erdogan, J.~R. Hershey, S.~Watanabe, and J.~Le~Roux, ``{Phase-sensitive and
  recognition-boosted speech separation using deep recurrent neural
  networks},'' in \emph{Proc. ICASSP}, 2015.

\bibitem{Han2020}
\BIBentryALTinterwordspacing
C.~Han, Y.~Luo, C.~Li, T.~Zhou, K.~Kinoshita, S.~Watanabe, M.~Delcroix,
  H.~Erdogan, J.~R. Hershey, N.~Mesgarani, and Z.~Chen, ``{Continuous speech
  separation using speaker inventory for long multi-talker recording},'' 2020.
  [Online]. Available: \url{http://arxiv.org/abs/2012.09727}
\BIBentrySTDinterwordspacing

\bibitem{Menne2019}
T.~Menne, I.~Sklyar, R.~Schl{\"{u}}ter, and H.~Ney, ``{Analysis of deep
  clustering as preprocessing for automatic speech recognition of sparsely
  overlapping speech},'' in \emph{Proc. Interspeech}, 2019.

\bibitem{ChenLibri2020}
Z.~Chen, T.~Yoshioka, L.~Lu, T.~Zhou, Z.~Meng, Y.~Luo, J.~Wu, X.~Xiao, and
  J.~Li, ``{Continuous speech separation: Dataset and analysis},'' in
  \emph{Proc. ICASSP}, 2020.

\bibitem{Li2021}
Y.~Li, Y.~Sun, K.~Horoshenkov, and S.~M. Naqvi, ``{Domain adaptation and
  autoencoder-based unsupervised speech enhancement},'' \emph{IEEE Transactions
  on Artificial Intelligence}, vol.~3, no.~1, pp. 43--52, 2021.

\bibitem{Li2021a}
C.~Li, Y.~Luo, C.~Han, J.~Li, T.~Yoshioka, T.~Zhou, M.~Delcroix, K.~Kinoshita,
  C.~Boeddeker, Y.~Qian, S.~Watanabe, and Z.~Chen, ``{Dual-path RNN for long
  recording speech separation},'' in \emph{Proc. IEEE SLT}, 2021.

\bibitem{Wichern2019}
G.~Wichern, J.~Antognini, M.~Flynn, L.~R. Zhu, E.~McQuinn, D.~Crow, E.~Manilow,
  and J.~Le~Roux, ``{Wham!: Extending speech separation to noisy
  environments},'' in \emph{Proc. Interspeech}, 2019.

\bibitem{Maclejewski2020}
M.~MacIejewski, G.~Wichern, E.~McQuinn, and J.~L. Roux, ``{WHAMR!: Noisy and
  reverberant single-channel speech separation},'' in \emph{Proc. ICASSP},
  2020.

\bibitem{Subakan2021a}
C.~Subakan, M.~Ravanelli, S.~Cornell, and F.~Grondin, ``{REAL-M: Towards speech
  separation on real mixtures},'' in \emph{Proc. ICASSP}, 2022.

\bibitem{Petermann2021}
D.~Petermann, G.~Wichern, Z.-Q. Wang, and J.~L. Roux, ``{The cocktail fork
  problem: Three-stem audio separation for real-world soundtracks},'' in
  \emph{Proc. ICASSP}, 2022.

\bibitem{Kadioglu2020}
B.~Kadioǧlu, M.~Horgan, X.~Liu, J.~Pons, D.~Darcy, and V.~Kumar, ``{An
  empirical study of Conv-TasNet},'' in \emph{Proc. ICASSP}, 2020.

\bibitem{cosentino2020}
\BIBentryALTinterwordspacing
J.~Cosentino, M.~Pariente, S.~Cornell, A.~Deleforge, and E.~Vincent,
  ``{LibriMix: An open-source dataset for generalizable speech separation},''
  2020. [Online]. Available: \url{http://arxiv.org/abs/2005.11262}
\BIBentrySTDinterwordspacing

\bibitem{Borsdorf2021}
M.~Borsdorf, C.~Xu, H.~Li, and T.~Schultz, ``{GlobalPhone mix-to-separate out
  of 2: A multilingual 2000 speakers mixtures database for speech
  separation},'' in \emph{Proc. Interspeech}, 2021.

\bibitem{Maciejewski2019}
M.~Maciejewski, G.~Sell, Y.~Fujita, L.~P. Garcia-Perera, S.~Watanabe, and
  S.~Khudanpur, ``{Analysis of robustness of deep single-channel speech
  separation using corpora constructed from multiple domains},'' in \emph{Proc.
  WASPAA}, 2019.

\bibitem{Tseng2003}
C.-y. Tseng, Y.-c. Cheng, W.-s. Lee, and F.-l. Huang, ``{Collecting Mandarin
  speech databases for prosody investigations},'' in \emph{Proc. O-COCOSDA},
  2003.

\bibitem{Liao2020}
Y.~F. Liao, C.~Y. Chang, H.~K. Tiun, H.~L. Su, H.~L. Khoo, J.~S. Tsay, L.~K.
  Tan, P.~Kang, T.~G. Thiann, U.~G. Iunn, J.~H. Yang, and C.~N. Liang,
  ``{Formosa speech recognition challenge 2020 and taiwanese across Taiwan
  corpus},'' in \emph{Proc. O-COCOSDA}, 2020.

\bibitem{Povey2011}
D.~Povey, O.~Glembek, N.~Goel, A.~Ghoshal, G.~Boulianne, L.~Burget, O.~Glembek,
  M.~Hannemann, P.~Motl{\'{i}}{\v{c}}ek, Y.~Qian, P.~Schwarz, J.~S. Silovsk´y,
  G.~Stemmer, and K.~V. Vesel´y, ``{The Kaldi speech recognition toolkit},''
  in \emph{Proc. IEEE ASRU}, 2011.

\bibitem{Luo2019}
Y.~Luo and N.~Mesgarani, ``{Conv-TasNet : Surpassing ideal time-frequency
  magnitude masking for speech separation},'' \emph{IEEE/ACM Trans. Audio,
  Speech, Lang. Process.}, vol.~27, no.~8, pp. 1256--1266, 2019.

\bibitem{Pariente2020}
M.~Pariente, S.~Cornell, J.~Cosentino, S.~Sivasankaran, E.~Tzinis,
  J.~Heitkaemper, M.~Olvera, F.~R. St{\"{o}}ter, M.~Hu, J.~M.
  Mart{\'{i}}n-Do{\~{n}}as, D.~Ditter, A.~Frank, A.~Deleforge, and E.~Vincent,
  ``{Asteroid: The PyTorch-based audio source separation toolkit for
  researchers},'' in \emph{Proc. Interspeech}, 2020.

\bibitem{Healy2021}
E.~W. Healy, E.~M. Johnson, M.~Delfarah, D.~S. Krishnagiri, V.~A. Sevich,
  H.~Taherian, and D.~Wang, ``{Deep learning based speaker separation and
  dereverberation can generalize across different languages to improve
  intelligibility},'' \emph{The Journal of the Acoustical Society of America},
  vol. 150, no.~4, pp. 2526--2538, 2021.

\bibitem{sivaraman2021}
A.~Sivaraman, S.~Wisdom, H.~Erdogan, and J.~R. Hershey, ``{Adapting speech
  separation to real-world meetings using mixture invariant training},'' in
  \emph{Proc. ICASSP}, 2022.

\end{thebibliography}

\end{document}